\documentclass[fleqn,twoside]{article}
\usepackage{espcrc2}
\usepackage{epsf}

\usepackage{graphicx}
\usepackage[figuresright]{rotating}

\newcommand{\AmS}{{\protect\the\textfont2
  A\kern-.1667em\lower.5ex\hbox{M}\kern-.125emS}}
\def\d{\delta}

\def\be{\begin{equation}}
\def\ee{\end{equation}}
\def\bea{\begin{eqnarray}}
\def\eea{\end{eqnarray}}

\title{Confinement made simple in the Coulomb gauge\thanks{Talk
                    presented by D.\ Zwanziger}}

\author{Attilio Cucchieri\address[Bras]{IFSC-USP, Caixa Postal 369,
                                    13560-970 S\~ao Carlos, SP,
                                    Brazil}\thanks{Research partially
        supported by FAPESP, Brazil (Project No.\ 00/05047-5).
        E-mail address: attilio@if.sc.usp.br}
        and
        Daniel Zwanziger\address{Department of Physics, NYU,
                              New York, USA}\thanks{Research partially
                              supported
                              by the National Science Foundation,
                              grant no.\ PHY-0099393.
                              E-mail address: daniel.zwanziger@nyu.edu
                              \vspace*{0.1cm}}}

\begin{document}

\begin{abstract}

In Gribov's scenario in Coulomb gauge, confinement of color charge is due to
a long-range instantaneous color-Coulomb potential $V(R)$. This may be
determined numerically from the instantaneous part of the gluon propagator
$D_{44, {\rm inst}} = V(R)\delta(t)$.  Confinement of gluons is reflected in
the vanishing at ${\bf k} = \bf{0}$ of the equal-time three-dimensionally
transverse would-be physical gluon propagator $D^{\rm tr}({\bf k})$.
We present exact analytic
results on $D_{44}$ and $D^{\rm tr}$ (which have also been investigated
numerically, A.\ Cucchieri, T.\ Mendes, and D.\ Zwanziger, this conference),
in particular the vanishing of $D^{\rm tr}({\bf k})$ at ${\bf k} = \bf{0}$,
and the determination
of the running coupling constant from $x_0 g^2({\bf k}) = {\bf k}^2D_{44,
{\rm inst}}$,
where $x_0 = 12N/(11N-2N_f)$.

\end{abstract}

\maketitle

	In QCD a rectangular Wilson loop $W(R, T)$ of dimension $R \times T$
has, asymptotically at large $T$, the form $W(R,T) \sim \exp[-T V_W(R)]$,
where $V_W(R)$ is the Wilson potential.  If dynamical quarks are
present, they are polarized from the vacuum, and $V_W(R)$ represents the
interaction energy of a pair of mesons at separation $R$.  In
this case $V_W(R)$ is not a color-confining potential, but rather
a QCD analog of the van der Waals potential between neutral atoms.  It
clearly cannot serve as an order parameter for confinement of color in
the presence of dynamical quarks, and we turn instead to gauge-dependent
quantities to characterize color confinement.

\setcounter{footnote}{0}

	A particularly simple confinement scenario \cite{gribov,coul}
is
available in the minimal Coulomb gauge\footnote{
The minimal lattice Coulomb gauge is obtained by first minimizing
$-\sum_{x,i=1}^3 {\rm Tr} {^gU}_{x,i}$ with respect to all local gauge
transformations $g(x)$, and then minimizing $-\sum_{x} {\rm Tr}
{^gU}_{x,4}$ with respect to all ${\bf x}$-independent but
$x_4$-dependent gauge transformations $g(x_4)$. This makes the 3-vector
potential $A_i$, for $i = 1,2,3$ transverse,
$\partial_i A_i = 0$, so $A_i =
A_i^{\rm tr}$. Moreover, the Coulomb gauge is the finite  limit of
renormalizable gauges \cite{recoul}.}. It attributes confinement
of color to the {\it enhancement} at long range of the color-Coulomb
potential $V(R)$.  This quantity is the instantaneous part of the 4-4
component of the gluon propagator,
$D_{\mu \nu}(x) \equiv  \langle g A_\mu(x) gA_\nu(0) \rangle$,
namely
$D_{44}({\bf x}, t) = V(|{\bf x}|) \d(t) + P({\bf x}, t)$.  The vacuum
polarization term $P({\bf x}, t)$ is less singular  than $\delta(t)$  at
$t = 0$.  At the same time, the disappearance of gluons from the physical
spectrum is manifested by the {\it suppression} at ${\bf k} = \bf{0}$ of
the propagator $D_{ij}({\bf k}, k_4)$ of 3-dimensionally transverse
would-be physical gluons.  This  qualitative behavior is clearly exhibited
in the numerical studies reported in
\cite{cuzwns}, \cite{fitgribov} and \cite{cmz} which
display the lattice equal-time propagators
$D^{\rm tr}({\bf k})$ and $D_{44}({\bf k})$.  Here
$D_{ij}({\bf k},k_4) =
(\delta_{ij}-\hat{k}_i\hat{k}_j)D^{\rm tr}({\bf k},k_4)$
designates the Fourier  transform of
$D_{ij}({\bf x},t)$, and its equal-time part is given by
$D^{\rm tr}({\bf k}) = (2\pi)^{-1} \int dk_4 D^{\rm tr}({\bf k}, k_4)$.

	We may identify $V(R)$ with the phenomenological potential that is the
starting point for QCD bound state calculations \cite{V},
\cite{szcz}, \cite{robertson} and \cite{szczsw}.
It is tempting to conjecture that the
color-Coulomb potential $V(R)$ is linearly rising at large $R$ (at least
when asymptotic freedom holds) and that this linear rise may serve as an
order parameter for color confinement even in the presence of dynamical
quarks~\cite{coul}.

	The identification of the phenomenological potential with
the quantity $V(R)$ that appears in the gluon propagator, a fundamental
quantity in the gauge theory, is made possible by several remarkable
properties of $V(R)$.  Note first that $A_4$ couples
universally to color charge, so a long range of the instantaneous part of
$D_{44}$ can confine all color charge.  Secondly one may show \cite{rgcoul}
that $V(R)$
is the expectation-value of the ${\bf A}$-dependent potential that appears
in the QCD Hamiltonian in Coulomb gauge,
$ V({\bf x - y}) =
\langle (M^{-1} (- \nabla^2)M^{-1})|_{\bf x,y} \rangle$,
where $M({\bf A}) = - {\bf \nabla} \cdot {\bf D}{\bf A})$ is the
3-dimensional Faddeev-Popov operator, and
${\bf D}({\bf A}) = {\bf \nabla} + {\bf A} \times$ is the gauge-covariant
derivative.   Finally, and most importantly,
$V(R)$ is a renormalization-group
invariant, and thus  scheme-independent, so it is independent of the
cut-off $\Lambda$ and of the renormalization mass $\mu$. This follows from
the non-renormalization of $gA_4$, as expressed by the identity
$g_{(0)}A_4^{(0)} = g_{(r)}A_4^{(r)}$, where $0$ and $r$ refer to
unrenormalized and renormalized quantities in the Coulomb gauge
\cite{coul}.  This identity has no direct analog in a Lorentz-covariant
gauge.  Because of the scheme-independence of $V(R)$, its
Fourier transform $\widetilde{V}({\bf k})$ provides a
scheme-independent definition the running
coupling constant of QCD,
${\bf k}^2\widetilde{V}({\bf k})
= x_0 \ g_{\rm coul}^2(|{\bf k}|)$, and of
$\alpha_s \equiv { {g^2({\bf k}/\Lambda_{\rm coul})} \over {4\pi}}$. Here
$x_0 = { {12N} \over {11N - 2N_f} }$, and $\Lambda_{\rm coul}$ is a finite
QCD mass scale \cite{rgcoul}.
{\it All} coefficients in the expansion of the
$\beta$-function
$\beta_{\rm coul}(g) \equiv |{\bf k}|
{ {\partial g_{\rm coul}} \over{\partial |{\bf k}|} }
= -(b_0g^3 + b_1g^5 +b_2g^7 + \dots)$, are scheme-independent \cite{rgcoul}.
One may determine $\alpha_s$ from a numerical
evaluation of the equal-time 2-point function $D_{44}$ in the minimal
Coulomb gauge \cite{cuzwns}.

A less intuitive but equally striking prediction concerns the
3-dimensionally transverse, equal-time would-be physical gluon propagator
$D^{\rm tr}({\bf k})$.
It was proven \cite{vanish}, as a consequence of the Gribov horizon
that, for infinite spatial lattice volume $L^3$, $D^{\rm tr}({\bf k})$
{\it vanishes} at ${\bf k} = \bf{0}$,
$\lim_{|{\bf k}| \to 0}D^{\rm tr}({\bf k}) = 0$.
This is in marked contrast to the free equal-time massless propagator,
$(2|{\bf k}|)^{-1} = (2\pi)^{-1} \int dk_4 ({\bf k}^2 + k_4^2)^{-1}$, that
{\it diverges} at ${\bf k} = \bf{0}$.  However the rate of approach of
$D^{\rm tr}({\bf k}, L)$ to 0, as a function of  ${\bf k}$ or the lattice
size $L$, was not established, and for this one must turn to numerical
studies.  Here it is essential to control effects of the finite-volume
$V = L^4$, at fixed $\beta = 2n/g_0^2$.  They are severe in the infrared
region, and the vanishing of $D^{\rm tr}({\bf k})$ at ${\bf k} = \bf{0}$
does not hold at finite $L$.  An extrapolation to infinite $L$ was done
in \cite{cuzwns}, \cite{fitgribov} and \cite{cmz},
and was found to be consistent with
$D^{\rm tr}({\bf 0}) = 0$.  It is also predicted that the gluon
propagator in Landau gauge $D(k)$ also vanishes at $k = 0$.  Numerical
studies of the gluon propagator in Landau gauge may be found in
\cite{cmz}, \cite{mandula}, \cite{gluonalf}, \cite{cuinfrared},
\cite{leinweber}, and~\cite{langfeldet}.

	Finally we recall that both the infrared enhancement of $D_{44}$
and the
infrared suppression of $D_{ij}$ may be understood as a result of the
restriction to the Gribov region in the minimal Coulomb gauge.  Indeed,
the proximity of the Gribov horizon in infrared directions suppresses the
infrared components of $A_i({\bf k})$ and thus of the gluon propagator
$D({\bf k}) = \langle |A({\bf k})|^2 \rangle$.  Secondly because the
Gribov region is the region where the Faddeev-Popov operator
$M({\bf A}) = - {\bf \nabla} \cdot {\bf D}(A)$ is positive, its boundary
occurs where $M({\bf A})$ has a (non-trivial) zero eigenvalue.  Entropy
then strongly favors high population density near the boundary, where
$M({\bf A})$ has a very small eigenvalue.  This enhances $V(R)$ by virtue
of the formula for $V(R)$ given above.

\end{document}